\documentclass[aps,prb,showpacs,twocolumn,superscriptaddress]{revtex4}
\usepackage{amsmath,graphics}
\usepackage[next]{inputenc}
\usepackage{graphicx}
\usepackage{hyperref}
\def\be{\begin{equation}}
\def\ee{\end{equation}}

\def\bi{\begin{itemize}}
\def\ei{\end{itemize}}
\def\bn{\begin{enumerate}}
\def\en{\end{enumerate}}
\def\bea{\begin{eqnarray}}
\def\eea{\end{eqnarray}}
\def\no{\nonumber}
\def\ba{\begin{array}}
\def\ea{\end{array}}
\def\bd{\begin{displaymath}}
\def\ed{\end{displaymath}}

\begin{document}

\title{Quantum Phase Transition in the One-Dimensional Extended Quantum Compass Model in a Transverse Field}
\author{R. Jafari}
\affiliation{Research Department, Nanosolar System Company (NSS), Zanjan 45158-65911, Iran}
\email[]{jafari@iasbs.ac.ir, jafari@nss.co.ir}
\affiliation{Department of Physics, Institute for Advanced
Studies in Basic Sciences (IASBS), Zanjan 45137-66731,Iran}

\begin{abstract}
Quantum phase transitions in the one-dimensional extended quantum compass model in transverse field are
studied by using the Jordan-Wigner transformation. This model is always gapful except at the critical surfaces where the energy gap disappears. We obtain the analytic expressions of all critical fields which drive quantum phase transitions. This model shows a rich phase diagram which includes spin-flop, strip antiferromagnetic and saturate ferromagnetic phases in addition to the phase with anti parallel ordering of spin $y$ component on odd bonds. However we study the universality and scaling properties of the transverse susceptibility and nearest-neighbor correlation functions derivatives in different regions to confirm the results obtained using the energy gap analysis.
\end{abstract}
\date{\today}


\maketitle

%%%%%%%%%%%%%%%%%%%%%%%%%%%%%%%%%%%%%%%%%%%%%%%%%%%%%%%%%%%%%%%%%%%%%
\section{Introduction \label{introduction}}

In the last decade, through the extensive experimental and theoretical works, the role
of orbital degree of freedom in determining the magnetic and transport properties
of transition-metal oxide materials has been recognized extensively \cite{Goodenough,Tokura,Wakabayashi}.
The complex interplay among the orbital, charge, spin, and lattice degrees of
freedom makes their phase diagrams extremely rich and leads various fascinating
physical phenomena. For instance, ferroelectricity, colossal magnetoresistance,
and charge ordering are the results of the orbital degeneracy in d-shell transition
metal oxides \cite{Cheong}.

A simplified model which described the nature of the orbital states in
the case of a twofold degeneracy is the Quantum Compass Model (QCM) \cite{Kugel}.
First, the model has been used to describe the Mott insulators with orbit degeneracies.
It depends on the lattice geometry and belongs to the low energy Hamiltonian originated from the
magnetic interactions in Mott-Hubbard systems with the strong spin-orbit coupling \cite{Jackeli}.
In QCM the orbital degrees of freedom are represented by pseudospin operators and
coupled anisotropically in such a way as to mimic the competition between orbital
orderings in different directions. For simplicity, the one-dimensional (1D) QCM, is
constructed by antiferromagnetic order of $X$ and $Y$ pseudospin components on odd and even bonds, respectively \cite{Brzezicki,Brzezicki2}. In addition, the 1D QCM is exactly the same
as the 1D reduced Kitaev model \cite{Feng}. Brzezicki \textit{et al}. obtained an exact solution
for the ground-state energy, reveals that the 1D QCM exhibits a first-order transition
between two disordered phases with opposite signs of certain local spin correlators.
Intriguingly, this first-order transition was found to be accompanied by a diverging correlation
length for spin correlations on one sublattice \cite{Brzezicki}.

Moreover, the extended version of the 1D QCM, obtained by introducing one more tunable parameter,
has been studied by Eriksson \textit{et al}. \cite{Eriksson}. They have identified four distinct ground
state phases, are separated by two intersecting transition lines. One of them defines a line
of second-order Ising-like transitions, while the other is a line of first-order transitions. The
point of intersection, where the first-order quantum phase transition identified by Brzezicki \textit{et al}
takes place, defines a multicritical point. They show that diverging of correlation length for certain
spin correlations, has recognized a natural explanation of the multicriticality of the transition point.
However, their results for the entanglement show that the only effect on the
ground state when going through the first order transitions is that a correlation function
for neighboring spins on odd bonds changes sign, without any effect on the entanglement
measures. First-order quantum phase transitions (QPT) are generally associated with a discontinuity in concurrence,
but accidental exceptions to this rule are possible.
To the best of our knowledge, the QCM in a transverse field has not been studied so far,
except on the first order transition line \cite{Sun}. They have shown that the energy gap does not disappear
in the presence of the transverse-field even in the thermodynamic limit.

In this paper, we study the 1D extended quantum compass model (EQCM) in a transverse  magnetic field.
The exact solution is obtained by using Jordan-Wigner (JW) transformation. We show that this model
reveal a rich phase diagram which includes quantum critical surfaces depending on exchange couplings. (We have been made aware that similar work is being performed by M. Motamedifar, S. Mahdavifar and S. Shayesteh Farjami using the exact diagonalization method, personal communication.)
Moreover, as we have shown in our resent work \cite{Jafari1}, because of nice scaling properties of correlation functions and transverse susceptibility (TS), phase transition can be captured from small systems with considerable accuracy without pre-assumed order parameters even for the cases where the pairwise entanglement is absent.
However we have exhibited that the divergence and scaling properties of two-body entanglement could be obtained by studying the correlation functions properties without direct calculation of the entanglement. So we will study the nearest neighbor correlation (NNC) functions  and TS of this model near a quantum critical points (QCP).

\section{Hamiltonian and Exact Solution\label{EQCMTF}}

Consider the Hamiltonian

\bea
\no
H=\sum_{n=1}^{N'}&[&J_{1}\sigma^{x}_{2n-1}\sigma^{x}_{2n}+
J_{2}\sigma^{y}_{2n-1}\sigma^{y}_{2n}+ L_{1}\sigma^{x}_{2n}\sigma^{x}_{2n+1}\\
\label{eq1}
&+&h(\sigma^{z}_{2n-1}+\sigma^{z}_{2n})].
\eea

where $J_{1}$ and $J_{2}$ are the odd bonds exchange couplings, $L_{1}$ is the even bond exchange coupling and $N=2N'$ is the number of spins.  We assume periodic boundary conditions.
The above Hamiltonian (Eq. (\ref{eq1})) can be exactly diagonalized by standard Jordan-Wigner transformation \cite{Jordan} as defined below,

\bea
\no
\sigma^{x}_{j}=b^{+}_{j}+b^{-}_{j},~~
\sigma^{y}_{j}=b^{+}_{j}+b^{-}_{j},~~
\sigma^{z}_{j}=2b^{+}_{j}b^{-}_{j}-1
\eea
\bea
\no
b^{+}_{j}=c^{\dag}_{j}~e^{i\pi\Sigma_{m=1}^{j-1}c^{\dag}_{m}c_{m}},~~
b^{-}_{j}=e^{-i\pi\Sigma_{m=1}^{j-1}c^{\dag}_{m}c_{m}}~c_{j}
\eea

which transforms spins into fermion operators $c_{j}$.

The crucial step is to define independent Majorana fermions \cite{Sengupta} at site $n$, $c_{n}^{q}\equiv c_{2n-1}$ and $c_{n}^{p}\equiv c_{2n}$. This can be regarded as quasiparticles' spin or as splitting the chain into bi-atomic
elementary cells \cite{Brzezicki2}.

Substituting for $\sigma^{x}_{j}$, $\sigma^{y}_{j}$ and $\sigma^{z}_{j}$ ($j=2n, 2n-1$) in terms of Majorana fermions with antiperiodic boundary condition (subspace with even number of fermions) followed by a Fourier transformation, Hamiltonian Eq. (\ref{eq1}) (apart from additive constant), can be written as

\bea
\label{eq2}
H^{+}=\sum_{k}\Big[Jc_{k}^{q\dag}c_{-k}^{p\dag}+Lc_{k}^{q\dag}c_{k}^{p}+
2h(c_{k}^{q\dag}c_{k}^{q}+c_{k}^{p\dag}c_{k}^{p})+h.c.\Big],
\eea

where $J=(J_{1}-J_{2})-L_{1}e^{ik}$, $L=(J_{1}+J_{2})+L_{1}e^{ik}$ and $k=\pm\frac{j\pi}{N'},~(j=1,3,\cdots,N'-1)$.\\

It should be pointed out that although the GS in periodic and antiperiodic boundary conditions are slightly different in the finite-size system, they are identical in the thermodynamic limit and the essential features in finite size are also not altered qualitatively.

Finally, diagonalization is completed by a four-dimensional Bogoliubov transformation connecting
$c_{k}^{q\dag},~c_{-k}^{q},~c_{k}^{p\dag},~c_{-k}^{p}$ and obtain two different kind of quasiparticles,

\bea
\label{eq3}
H=\sum_{k}\Big[E^{q}_{k}(\gamma_{k}^{q\dag}\gamma_{k}^{q}-\frac{1}{2})+
E^{p}_{k}(\gamma_{k}^{p\dag}\gamma_{k}^{p}-\frac{1}{2})\Big],
\eea

where $E^{q}_{k}=\sqrt{2(a+c)}$ and $E^{p}_{k}=\sqrt{2(a-c)}$, $c=\sqrt{a^{2}-b}$ in which

\bea
\no
a&=&2h^{2}+J_{1}^{2}+J_{2}^{2}+L_{1}^{2}+2L_{1}J_{2}\cos k,\\
\no
b&=&4[(J_{1}J_{2}-h^{2})^{2}+2J_{1}L_{2}(J_{1}J_{2}-h^{2})\cos k+J_{1}^{2}L_{1}^{2}].
\eea

The ground state ($E_{G}$) and the lowest excited state ($E_{E}$) energies are obtained from Eq.(\ref{eq3}),

\bea
\no
E_{G}=-\frac{1}{2}\sum_{k}(E^{q}_{k}+E^{p}_{k}),~~E_{E}=-\frac{1}{2}\sum_{k}(E^{q}_{k}-E^{p}_{k}),
\eea

where could be written as a function of $a$ and $b$,

\bea
\label{eq4}
E_{G}=-2\sum_{k>0}\sqrt{a+\sqrt{b}},~~E_{E}=-2\sum_{k>0}\sqrt{a-\sqrt{b}}
\eea

It is clear the ground state is separated from the lowest energy pseudospin excitation by a pseudospin
gap $\Delta=|E_{E}-E_{G}|$,	which vanishes at $h_{0}=\sqrt{J_{1}(J_{2}+L_{1})}$ and $h_{\pi}=\sqrt{J_{1}(J_{2}-L_{1})}$ in the thermodynamic limit.

It should be stressed here that the exact spectrum and the pseudospin
gap are the same as that obtained by Eriksson \textit{et al}. for $h=0$ using a different method \cite{Eriksson}.

So, the quantum phase transition (QPT) which could be driven by the transverse-field, depending on exchange couplings, occurs at $h_{0}$ and $h_{\pi}$.

\section{Phase Diagram\label{PD}}

The complete phase diagram of the extended compass model without transverse magnetic field has been reported in
Refs. [\onlinecite{Eriksson}] and [\onlinecite{Mahdavifar}]. They have shown that the first-order transition occurs at multicritical point where a line of first-order transition ($J_{1}/L_{1}=0$) meets with a line of second order transition ($J_{2}/L_{1}=1$). Also, There are four gapped phases in the exchange couplings' space,

\begin{itemize}
  \item (I) $J_{1}>0,~J_{2}<1$: In this region the ground state is in the Ne\'{e}l phase along the $x$ axis.
  \item (II) $J_{1}>0,~J_{2}>1$: In this case there is antiparallel ordering of spin $y$ component on odd bonds.
  \item (III) $J_{1}<0,~J_{2}>1$: In this case there is parallel ordering of spin $y$ component on odd bonds.
  \item (IV) $J_{1}<0,~J_{2}<1$: In this region the ground state is in the strip antiferromagnetic (SAF) phase.
 \end{itemize}

\begin{figure}
\begin{center}
\includegraphics[width=8cm]{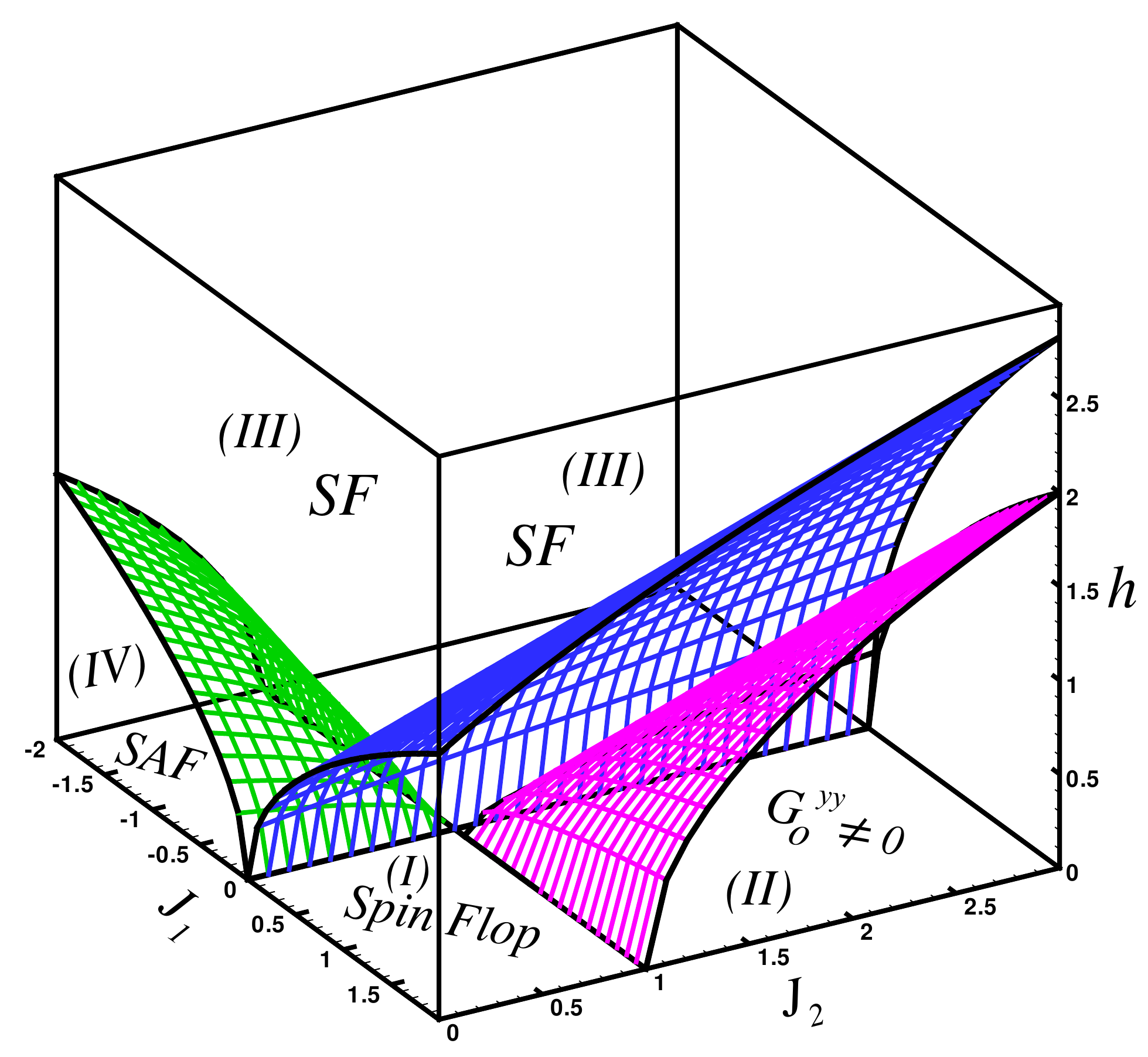}
\caption{(Color online) Phase diagram of the extended compass model in the transverse magnetic field.
For $J_{1}>0, J_{2}<1$, the front and top side of the purple convex surface is spin-flop phase (I) and the back side ($J_{1}>0, J_{2}>1$) specified by antiparallel ordered of spin $y$ component (II).
The  blue checkerboard pattern represents the boundary between spin-flop phase (I) and saturate ferromagnetic phase (III). In the case of $J_{1}<0, J_{2}<1$ there are two phases, the strip antiferromagnetic phase (IV) which exists below the green convex surface and the saturate ferromagnetic phase (III) which is above it. For $J_{1}<0, J_{2}>1$ saturate ferromagnetic phase (III) attends the phase diagram.  The first-order transition appears just at $J_{1}=0, h=0$ line.} \label{fig1}
\end{center}
\end{figure}

Phase diagram of extended quantum compass model in transverse field has been shown in Fig. (\ref{fig1}).
Depending on exchange couplings, the transverse field could drives the phase transition at $h_{0}$ and $h_{\pi}$ where the energy gap vanishes (For simplicity we take $L_{1}=1$).

%%%%%%%%%%%%%%%%%%%%%%  Fig.1   %%%%%%%%%%%%%%%%%%%%%%%
%%%%%%%%%%%%%%%%%%%%%%%%%%%%%%%%%%%%%%%%%%%%%%%%%%%%%%%

\begin{figure}
\begin{center}
\includegraphics[width=8cm]{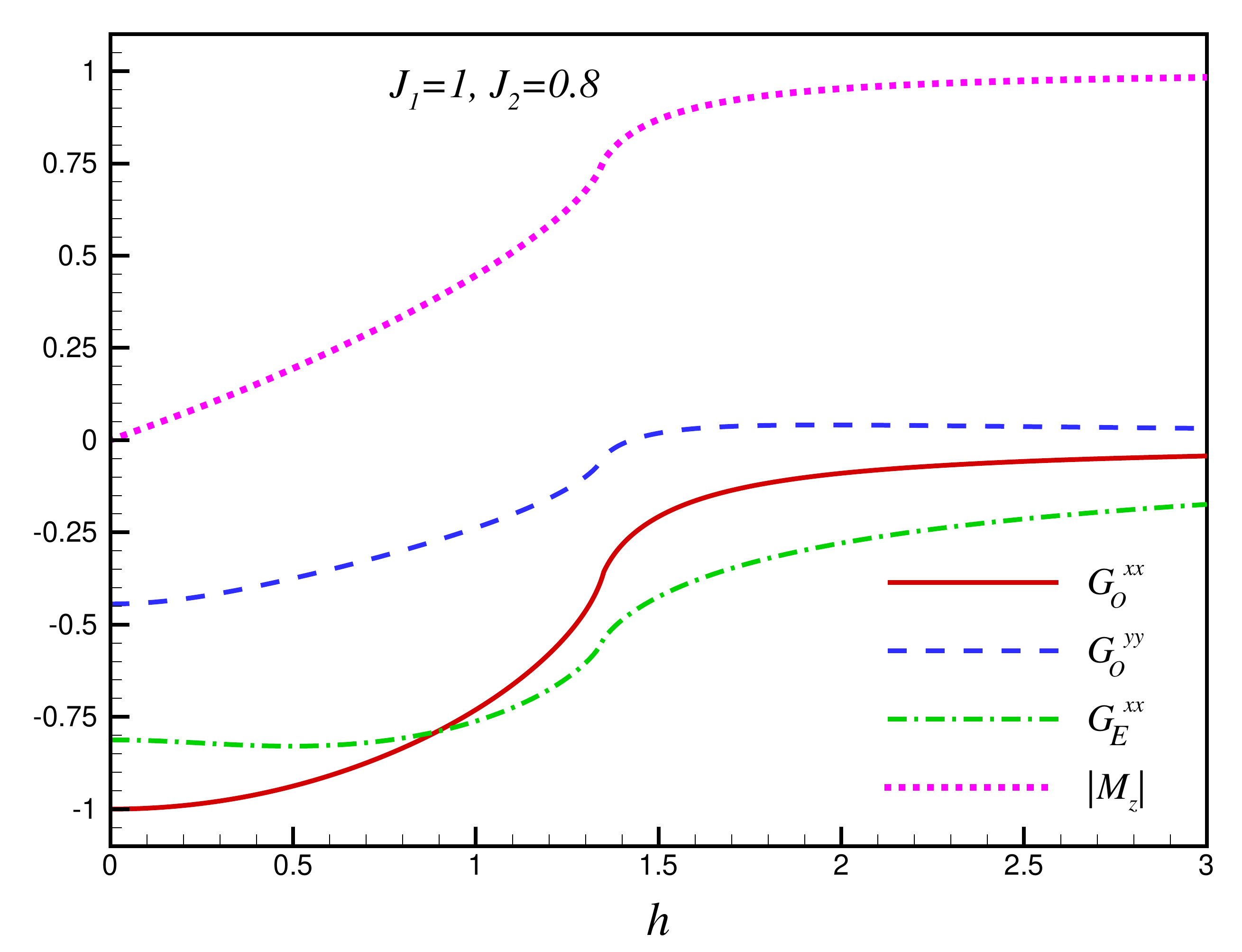}
\caption{(Color online) The transverse magnetization and different components of nearest-neighbor spin correlation
functions on even and odd bond for $J_{1}=1, J_{2}=0.8$.} \label{fig2}
\end{center}
\end{figure}

Fig. (\ref{fig2}) shows the absolute value of transverse magnetization (TM) and NNC functions on odd and even bonds for infinite system size in region (I) ($J_{1}=1, J_{2}=0.8$).
In this region tuning the magnetic field dictated the system
to fall into saturated ferromagnetic (SF) phase. The spin flop-SF phase transition occurs
at $h_{c}=h_{0}$ (blue checkerboard curved plane in Fig. (\ref{fig1})) which under this surface ground state is in the spin flop phase (the Ne\'{e}l ordered along the axis where is perpendicular to magnetic field is called spin flop). It is seen in Fig. (\ref{fig2}) that the onset of magnetic field sets up the TM ($M_{z}$) immediately and continuously increases with increase in $h$ to saturate value ($|M_{z}|=1$). However, the antiparallel ordered of spin $x$ and $y$ components on odd ($G^{xx}_{o}, G^{yy}_{o}$) and even ($G^{xx}_{E}$) bonds reduces by increasing the magnetic field and goes to zero for $h\rightarrow\infty$.

%%%%%%%%%%%%%%%%%%%%%%  Fig.2   %%%%%%%%%%%%%%%%%%%%%%%
%%%%%%%%%%%%%%%%%%%%%%%%%%%%%%%%%%%%%%%%%%%%%%%%%%%%%%%

%%%%%%%%%%%%%%%%%%%%%%  Fig.3   %%%%%%%%%%%%%%%%%%%%%%%
\begin{figure}[b]
\begin{center}
\includegraphics[width=8cm]{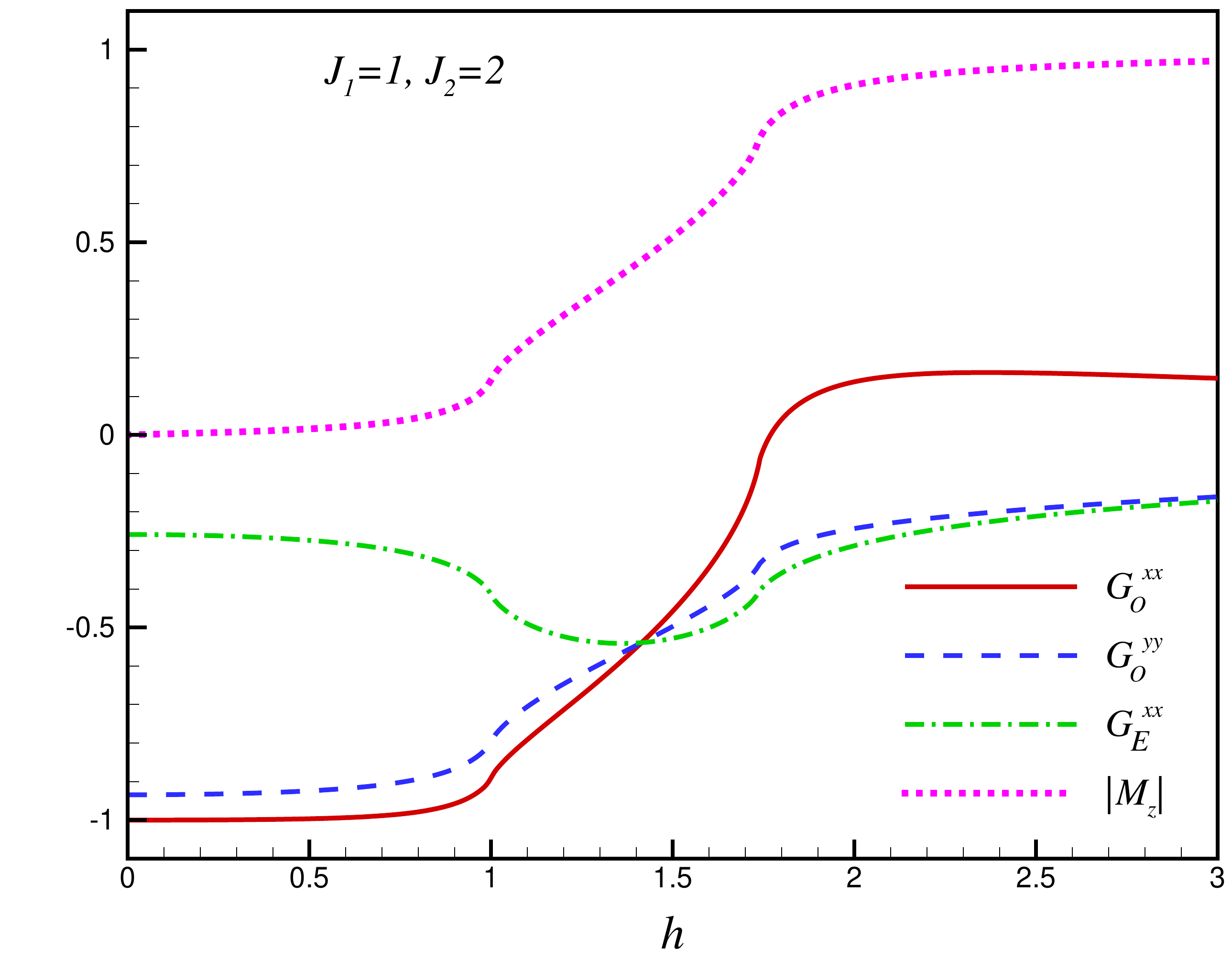}
\caption{(Color online) The TM and NNC
functions of spin components for $J_{1}=1, J_{2}=2$.} \label{fig3}
\end{center}
\end{figure}
%%%%%%%%%%%%%%%%%%%%%%%%%%%%%%%%%%%%%%%%%%%%%%%%%%%%%%%

In  region (II) the gap decreases by increasing the magnetic field and goes to zero at the lower critical field $h_{c_{1}}=h_{\pi}$ (purple checkerboard curved plane in Fig. (\ref{fig1})) and beyond this critical field the energy gap immediately appears with increase of the magnetic field. This process continues until the upper critical field $h_{c_{2}}=h_{0}$ (blue checkerboard curved plane in Fig. (\ref{fig1})) at which the energy gap vanishes and becomes once again gapped upon the $h_{c_{2}}$. Fig. (\ref{fig3}) shows the TM and NNC functions versus the magnetic field in the region (II) ($J_{1}=1, J_{2}=2$). It manifests that under the lower critical field ($h<h_{c_{1}}$) the antiparallel ordered of spin $x$ and $y$ components on odd and even bonds stay quite unchanged. However the TM is zero for $h<h_{c_{1}}$. So the ground state's antiparallel ordering of spin $y$ component remains unchanged under $h_{c_{1}}$.
Beyond the $h_{c_{1}}$ the TM and NNC functions undergo a strong qualitative
change and antiparallel ordered on odd and even bonds tends to zero as the magnetic field increases. Increasing the magnetic field saturates the TM and disappears the antiparallel ordered spin $x$ component on odd bound at $h_{c_{2}}=h_{0}$, while the antiparallel ordered spin $y$ component on odd bound and antiparallel ordered spin $x$ component on even bond have a nonzero values and tend to zero for $h\rightarrow\infty$.

%%%%%%%%%%%%%%%%%%%%%%  Fig.4   %%%%%%%%%%%%%%%%%%%%%%%
\begin{figure}[t]
\begin{center}
\includegraphics[width=8cm]{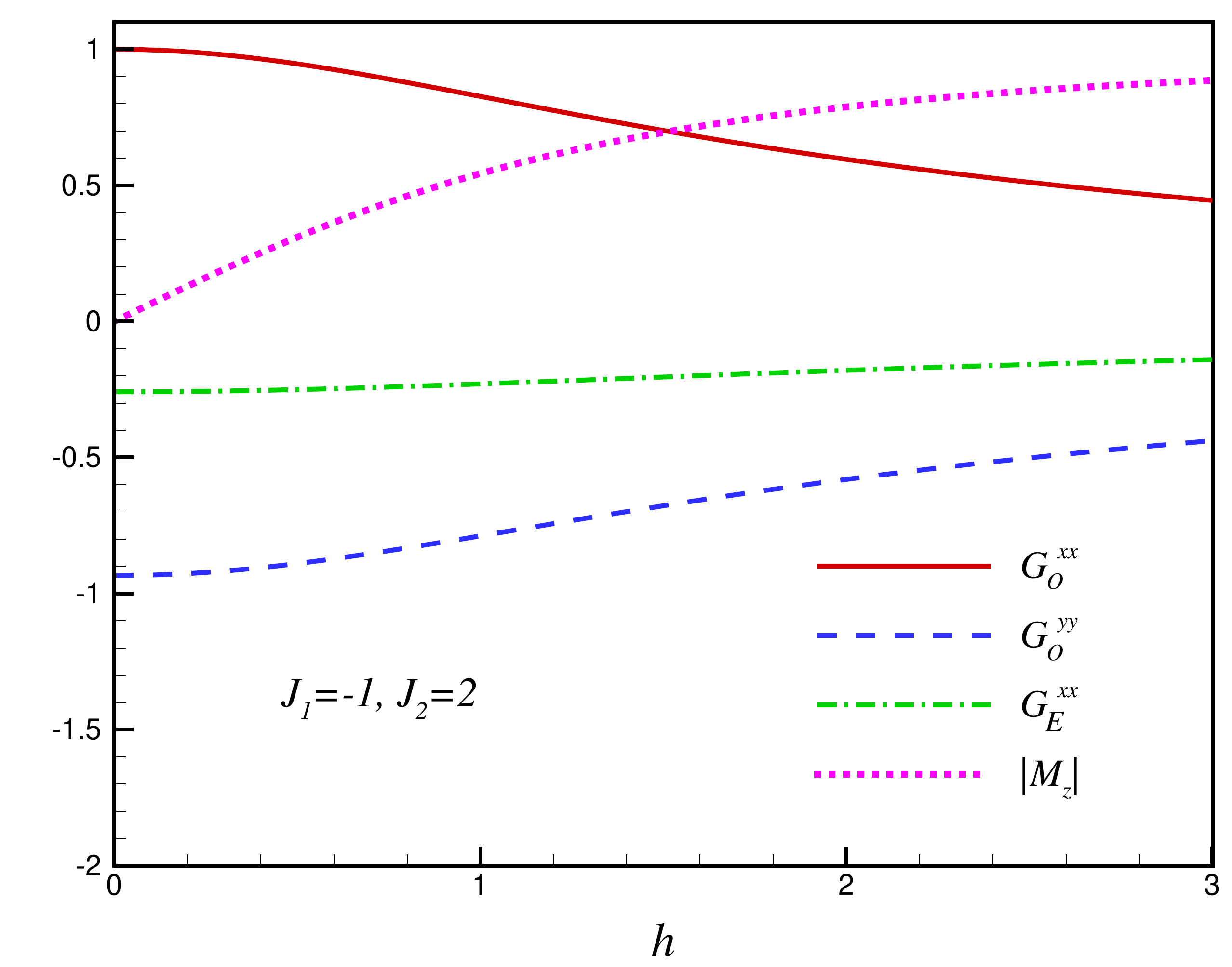}
\caption{(Color online) The TM and different components of nearest-neighbor spin correlation
functions on even and odd bond in region (III) ($J_{1}=-1, J_{2}=2$).} \label{fig4}
\end{center}
\end{figure}
%%%%%%%%%%%%%%%%%%%%%%%%%%%%%%%%%%%%%%%%%%%%%%%%%%%%%%%

A surprising result occurs
in the intermediate region of the magnetic fields $h_{c_{1}}<h<h_{c_{2}}$ where increasing the magnetic field enhances the antiparallel ordered spin $x$ component on even bond up to a maximum and then decreases
gradually, while decreases the other antiparallel ordered. So we predict that the gapped spin-flop phase exists in the intermediate values of the transverse magnetic field $h_{c_{1}}<h<h_{c_{2}}$. In other words, in the region (II), the magnetic field destroys the ground state's antiparallel ordering of spin $y$ component on even bond at $h_{c_{1}}$ and sticks the system in the spin-flop phase upon the $h_{c_{1}}$. The spin-flop-SF transition occurs beyond $h_{c_{2}}$.

The TM and NNC functions have been depicted in Fig. (\ref{fig4}) for region (III) ($J_{1}=-1, J_{2}=2$). In this case the system is in the SF phase and phase transition dose not occur by tuning the magnetic field even in the strong magnetic field.

%%%%%%%%%%%%%%%%%%%%%%  Fig.5   %%%%%%%%%%%%%%%%%%%%%%%
\begin{figure}[t]
\begin{center}
\includegraphics[width=8cm]{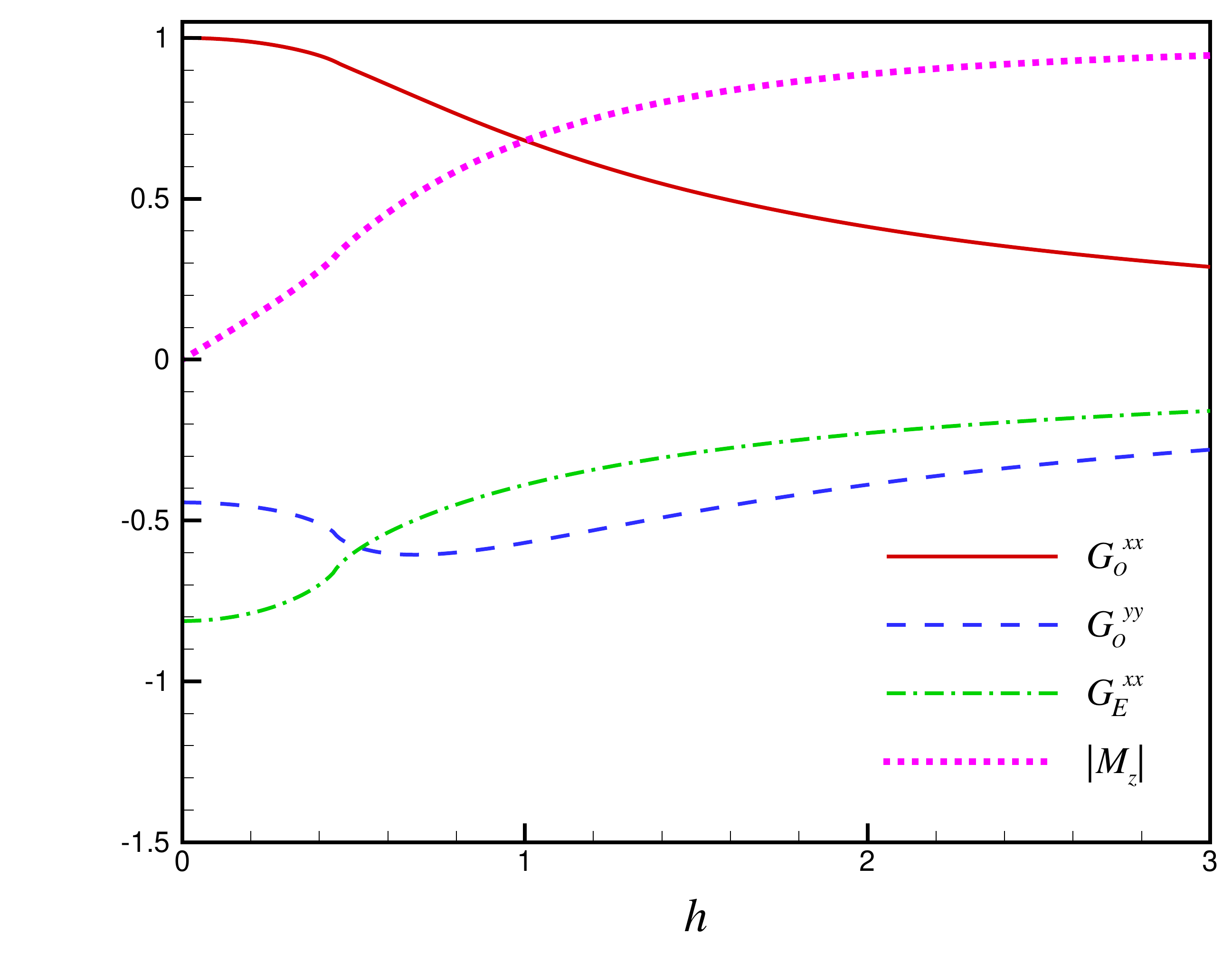}
\caption{(Color online) The TM and NNC
functions of spin components for $J_{1}=-1, J_{2}=0.8$.} \label{fig5}
\end{center}
\end{figure}
%%%%%%%%%%%%%%%%%%%%%%%%%%%%%%%%%%%%%%%%%%%%%%%%%%%%%%%

Fig. (\ref{fig5}) shows the TM and NNC functions in the region (IV) ($J_{1}=-1, J_{2}=0.8$). This region include two gapped phases, SAF and SF where separated from one another at the critical point $h_{c}=h_{\pi}$ (green checkerboard curved plane in Fig. (\ref{fig1})).

\begin{figure}[t]
\begin{center}
\includegraphics[width=9cm]{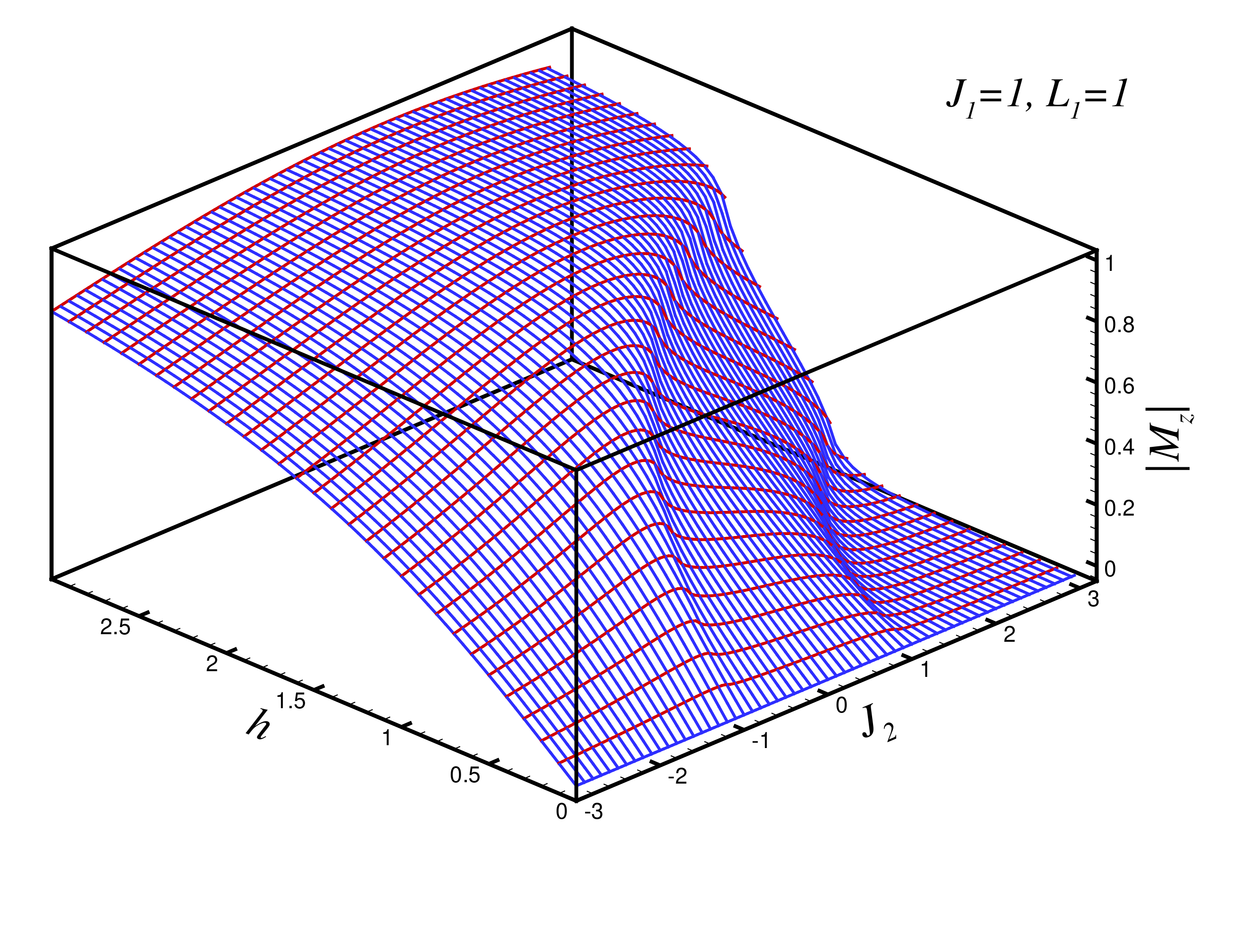}
\caption{(Color online) The three-dimensional panorama of TM for $J_{1}=1$.} \label{fig6}
\end{center}
\end{figure}

The three-dimensional panorama of TM with respect to $J_{2}$ and $h$ has been plotted in
Fig. (\ref{fig6}) for $J_{1}=1$. The two critical line ($h_{0}(J_{2},h)$ and $h_{\pi}(J_{2},h)$) at which the energy gap vanishes can be described by two assumed lines on the two convex part of the surface in Fig. (\ref{fig5}).

%%%%%%%%%%%%%%%%%%%%%%  Fig.6   %%%%%%%%%%%%%%%%%%%%%%%
%%%%%%%%%%%%%%%%%%%%%%%%%%%%%%%%%%%%%%%%%%%%%%%%%%%%%%%

\section{Universality and scaling of Correlation Functions\label{USCF}}

The nonanalytic behavior in some physical quantity is a feature of second-order quantum phase transition. It is also
accompanied by a scaling behavior since the correlation length diverges and there is no characteristic length scale in
the system at the critical point. As we previously mentioned, correlation functions and TS show the universality and scaling around the QCP and could capture QCP. However, studying the NNC functions behaviors could reveal the scaling and universality of entanglement near the QCP. So, in this section we will study the behavior of NNC functions derivative with respect to the magnetic field and TS to confirm the previous results.

%%%%%%%%%%%%%%%%%%%%%%  Fig.7   %%%%%%%%%%%%%%%%%%%%%%%
\begin{figure}[t]
\begin{center}
\includegraphics[width=8cm]{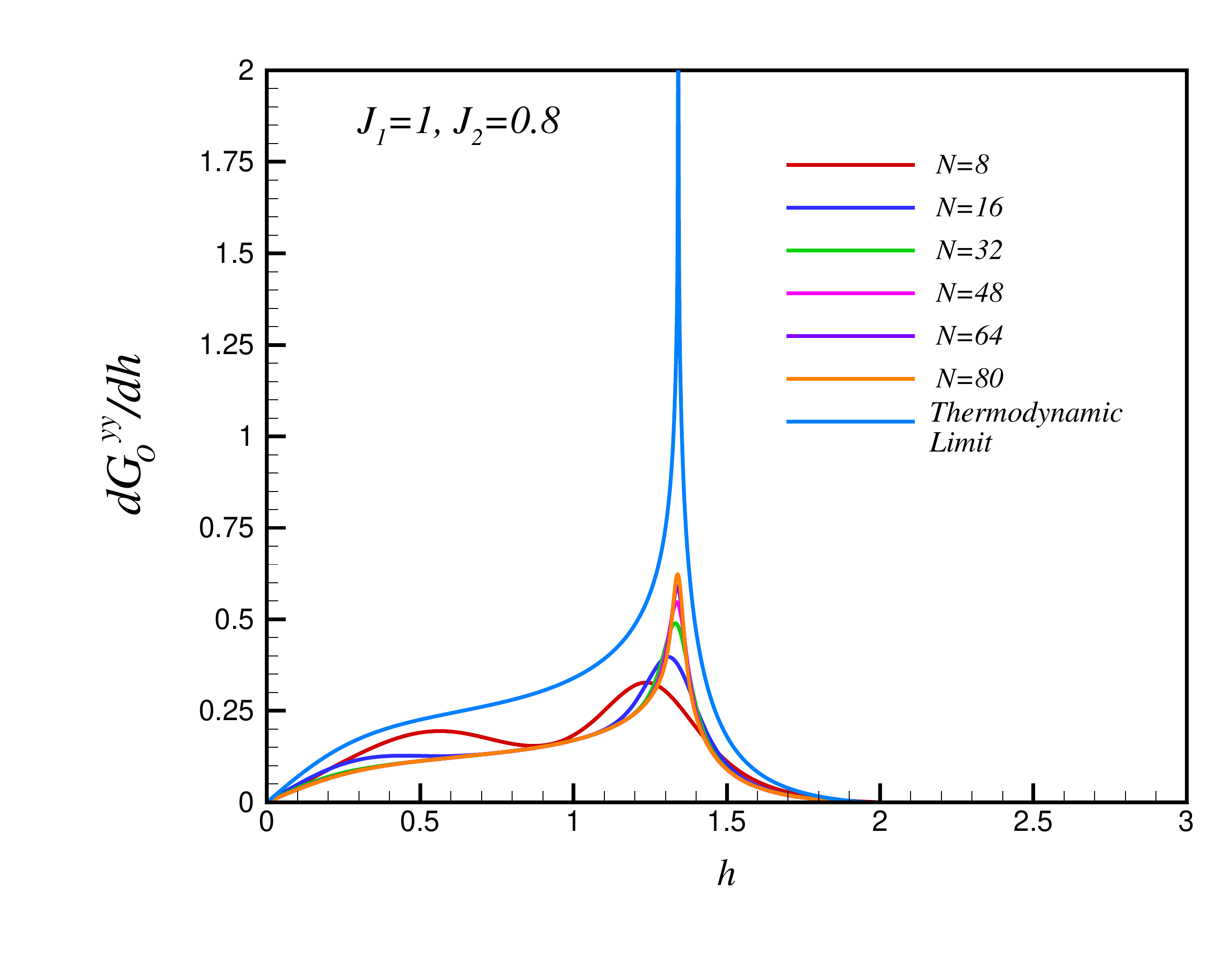}
\caption{(Color online) Evolution of the $\frac{dG^{yy}}{dh}$ versus $h$ for
different system sizes in region (I) for $J_{1}=1, J_{2}=0.8$.} \label{fig7}
\end{center}
\end{figure}
%%%%%%%%%%%%%%%%%%%%%%%%%%%%%%%%%%%%%%%%%%%%%%%%%%%%%%%

\begin{figure}
\begin{center}
\includegraphics[width=8cm]{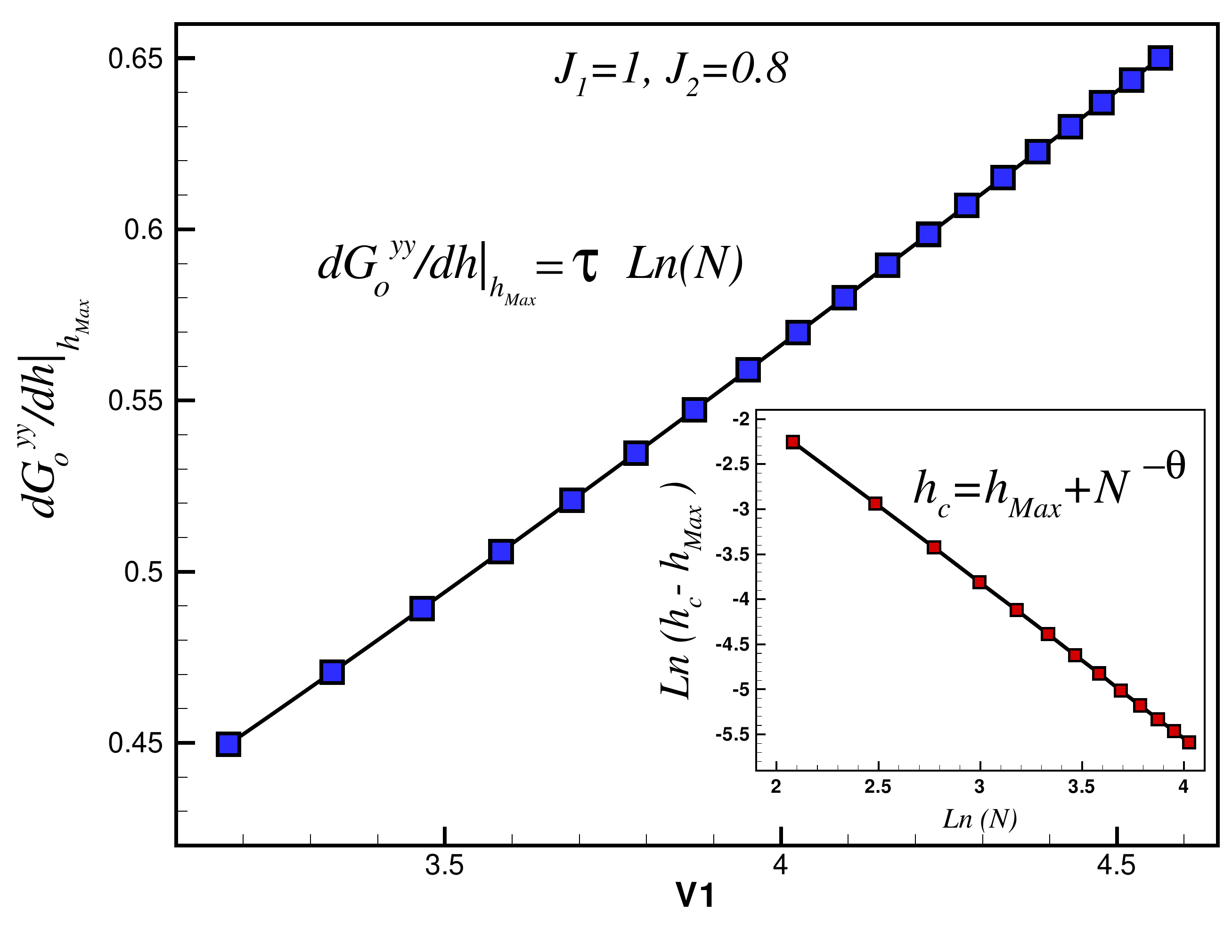}
\caption{(Color online) Scaling of the maximum of $\frac{dG^{yy}}{dh}$ for
systems of various sizes. Inset: Scaling of the position ($h_{Max}$) of $\frac{dG^{yy}}{dh}$
for different-length chains.} \label{fig8}
\end{center}
\end{figure}

In Fig. (\ref{fig7}) the derivative of $G_{o}^{yy}$ with respect to the magnetic field has been shown for different system sizes in region (I) ($J_{1}=1, J_{2}=2$). For infinite lattice size $dG_{o}^{yy}/dh$ diverges as the critical point is touched, while there is no divergence for finite lattice sizes. As the size of system becomes large, the derivative of NNC functions tends to diverge close to the critical point. More information can be obtained when the maximum values of each plot and their positions are analyzed. The position of the maximum ($h_{Max}$) of $dG_{o}^{yy}/dh$ tends toward the critical point like $h_{Max}=h_{c}-N^{-\theta}$ ($\theta=1.72\pm0.03$) which has been plotted in the inset of Fig. (\ref{fig8}).

%%%%%%%%%%%%%%%%%%%%%%  Fig.8   %%%%%%%%%%%%%%%%%%%%%%%
%%%%%%%%%%%%%%%%%%%%%%%%%%%%%%%%%%%%%%%%%%%%%%%%%%%%%%%

\begin{figure}[t]
\begin{center}
\includegraphics[width=8cm]{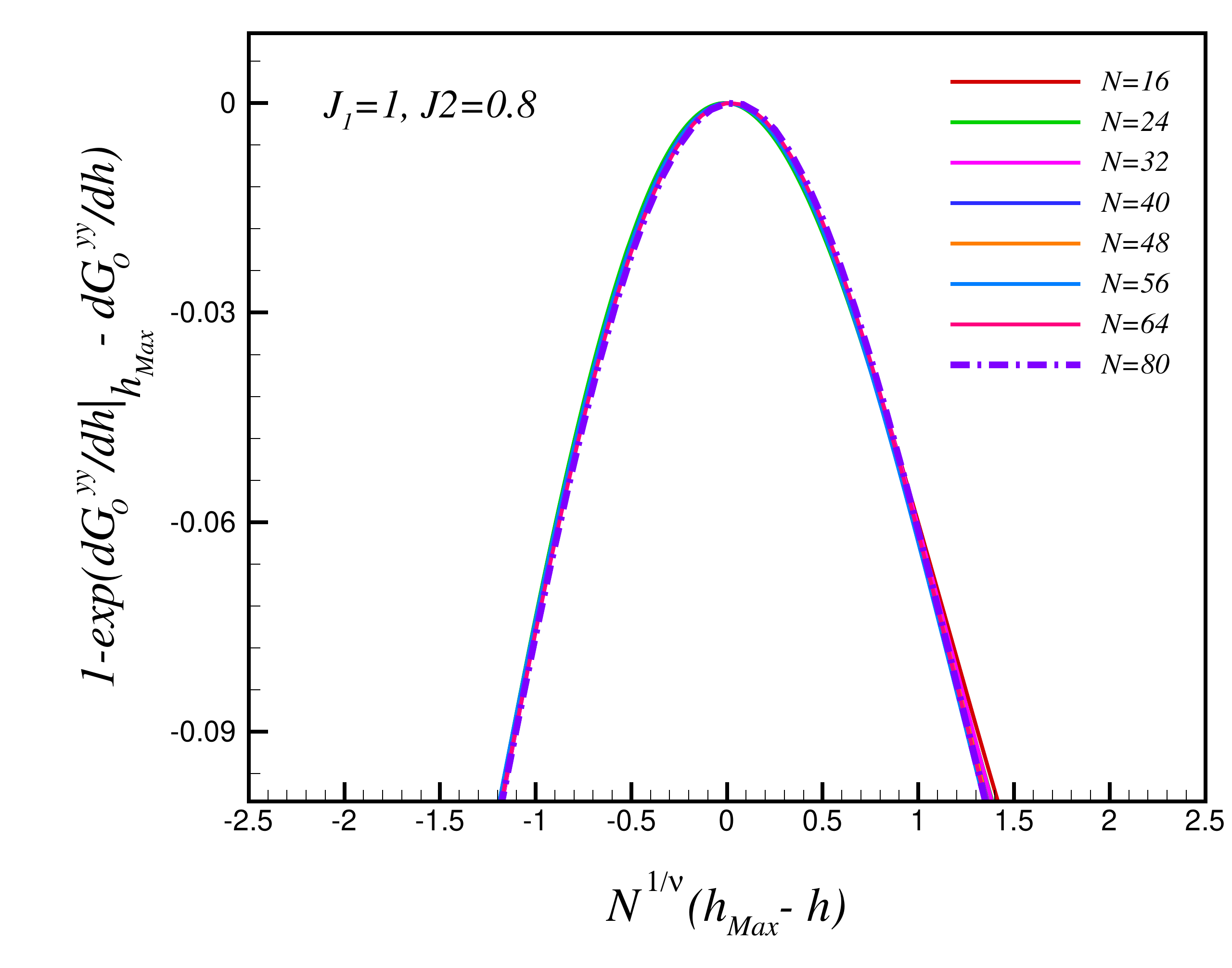}
\caption{(Color online) Finite-size scaling of $dG^{yy}/dh$
for different lattice sizes. The curves which correspond to
different system sizes clearly collapse on a single curve.} \label{fig9}
\end{center}
\end{figure}

Moreover, we have derived the scaling behavior of $|dG_{o}^{yy}/dh|_{h_{Max}}$ versus N. This has been plotted in  Fig. (\ref{fig8}) which shows a linear behavior of $|dG_{o}^{yy}/dh|_{h_{Max}}$ versus $\ln(N)$. The scaling behavior is $|dG^{yy}_{o}/dh|_{h_{Max}}=\tau\ln(N)$ with $\tau=0.15\pm0.01$.
To study the scaling behavior of $G_{o}^{yy}$ around the critical point, we perform finite-scaling analysis,
since the maximum value of derivative of $G_{o}^{yy}$ scales logarithmic. According to the scaling ansatz, the rescaled derivative of $G_{o}^{yy}$ around its maximum value $h_{Max}$ is just a function of rescaled driving parameter
such as
\bea
\label{eq5}
\frac{dG_{o}^{yy}}{dh}-\frac{dG_{o}^{yy}}{dh}|_{h_{Max}}\sim F(N^{1/\nu}(h-h_{Max})),
\eea
where $F(x)$ is a universal function. The manifestation of the finite-size scaling is shown in Fig. (\ref{fig9}). It is clear that the different curves which are resemblance of various system sizes collapse to a single universal curve.
Our result shows that $\nu=1\pm0.001$ is exactly correspond to the correlation length exponent of Ising model in transverse field ($\nu=1$).
%%%%%%%%%%%%%%%%%%%%%%  Fig.9   %%%%%%%%%%%%%%%%%%%%%%%
%%%%%%%%%%%%%%%%%%%%%%%%%%%%%%%%%%%%%%%%%%%%%%%%%%%%%%%

A similar analysis can be carried on $G_{o}^{xx}$, $G_{E}^{xx}$ and TS ($\chi^{z}$).
Our calculations show that the non-analytic and scaling behavior of
NN correlation functions are the same as TS does.
It is important to mention that NNC functions and TS show the logarithmic divergence near the QCP. Our results is
different from the reported result in Ref. [\onlinecite{Sun}]. They have found that TS shows a power-law behavior close to the QCP with the exponent $\gamma=1.78\pm0.05$. Since $h$ is analogous to temperature in classical systems, we expect $h\chi_{z}$ to be equivalent to the specific heat in the $2D$ classical Ising model ($J_{2}=0$). So the reported exponent in Ref. [\onlinecite{Sun}] belongs to susceptibility in the $x$-direction not to TS.

%%%%%%%%%%%%%%%%%%%%%%  Fig.10   %%%%%%%%%%%%%%%%%%%%%%%
\begin{figure}
\begin{center}
\includegraphics[width=8cm]{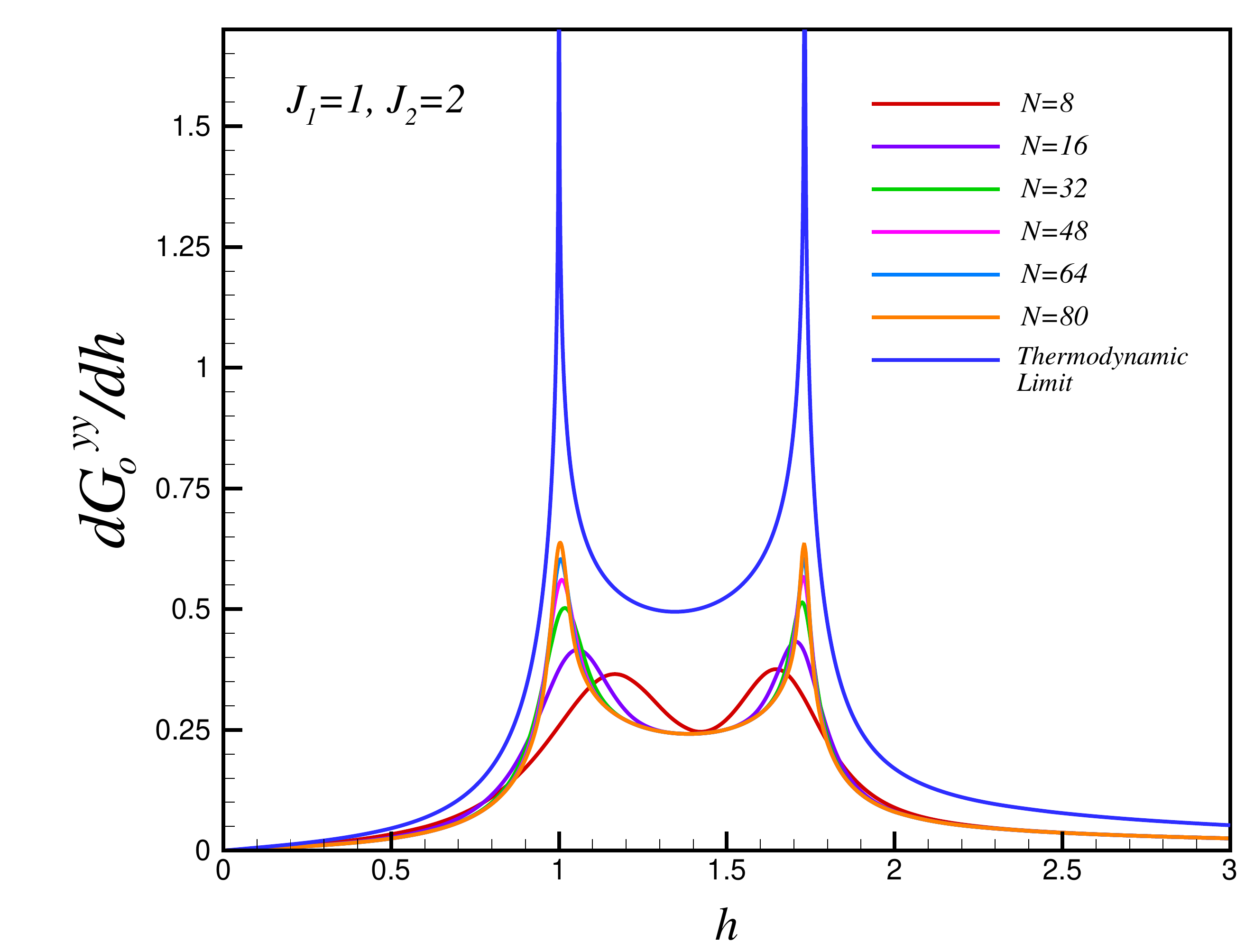}
\caption{(Color online) The first order derivative of $dG^{yy}/dh$ as a function
of $h$ for various system size in region (II) ($J_{1}=1, J_{2}=2$).} \label{fig10}
\end{center}
\end{figure}
%%%%%%%%%%%%%%%%%%%%%%%%%%%%%%%%%%%%%%%%%%%%%%%%%%%%%%%

\begin{figure}
\begin{center}
\includegraphics[width=8cm]{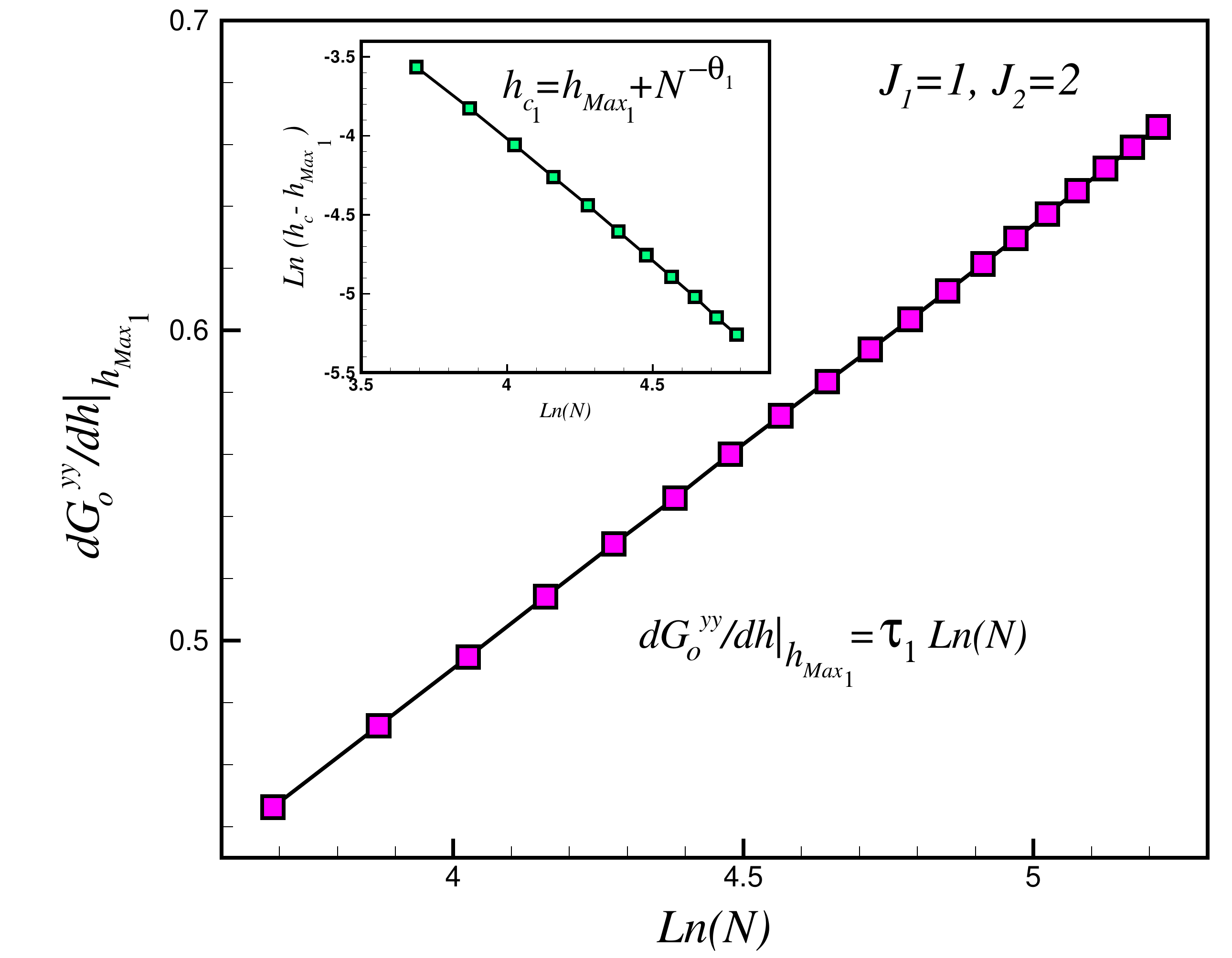}
\caption{(Color online) The scaling behavior of the first maximum point of $\frac{dG^{yy}}{dh}$ for
different-length chain in region (II). Inset: Scaling of the position ($h_{Max_{1}}$) of $\frac{dG^{yy}}{dh}$
for different-length chains ($J_{1}=1, J_{2}=2$).} \label{fig11}
\end{center}
\end{figure}

We have plotted the derivative of $G_{o}^{yy}$ in region (II) ($J_{1}=1, J_{2}=2$)
versus $h$ in Fig. (\ref{fig10}) for different lattice sizes which shows the singular behavior as the size of the system becomes large. As it manifests the divergences of $dG_{o}^{yy}/dh$ occur at $h_{c_{1}}=1$ and $h_{c_{2}}=\sqrt{3}$ where exactly correspond to the critical points that obtained using the energy gap analysis ($h_{c_{1}}=h_{\pi}, h_{c_{2}}=h_{0}$). A more detailed analysis manifest the linear behavior of $\frac{dG^{yy}_{o}}{dh}$ at the first maximum point ($h_{Max_{1}}$) versus $\ln(N)$ where has been plotted in Fig. (\ref{fig11}). The exponent for this behavior is $\tau_{1}=0.14\pm0.02$.

%%%%%%%%%%%%%%%%%%%%%%  Fig.11   %%%%%%%%%%%%%%%%%%%%%%%
%%%%%%%%%%%%%%%%%%%%%%%%%%%%%%%%%%%%%%%%%%%%%%%%%%%%%%%

%%%%%%%%%%%%%%%%%%%%%%  Fig.12   %%%%%%%%%%%%%%%%%%%%%%%
\begin{figure}
\begin{center}
\includegraphics[width=8cm]{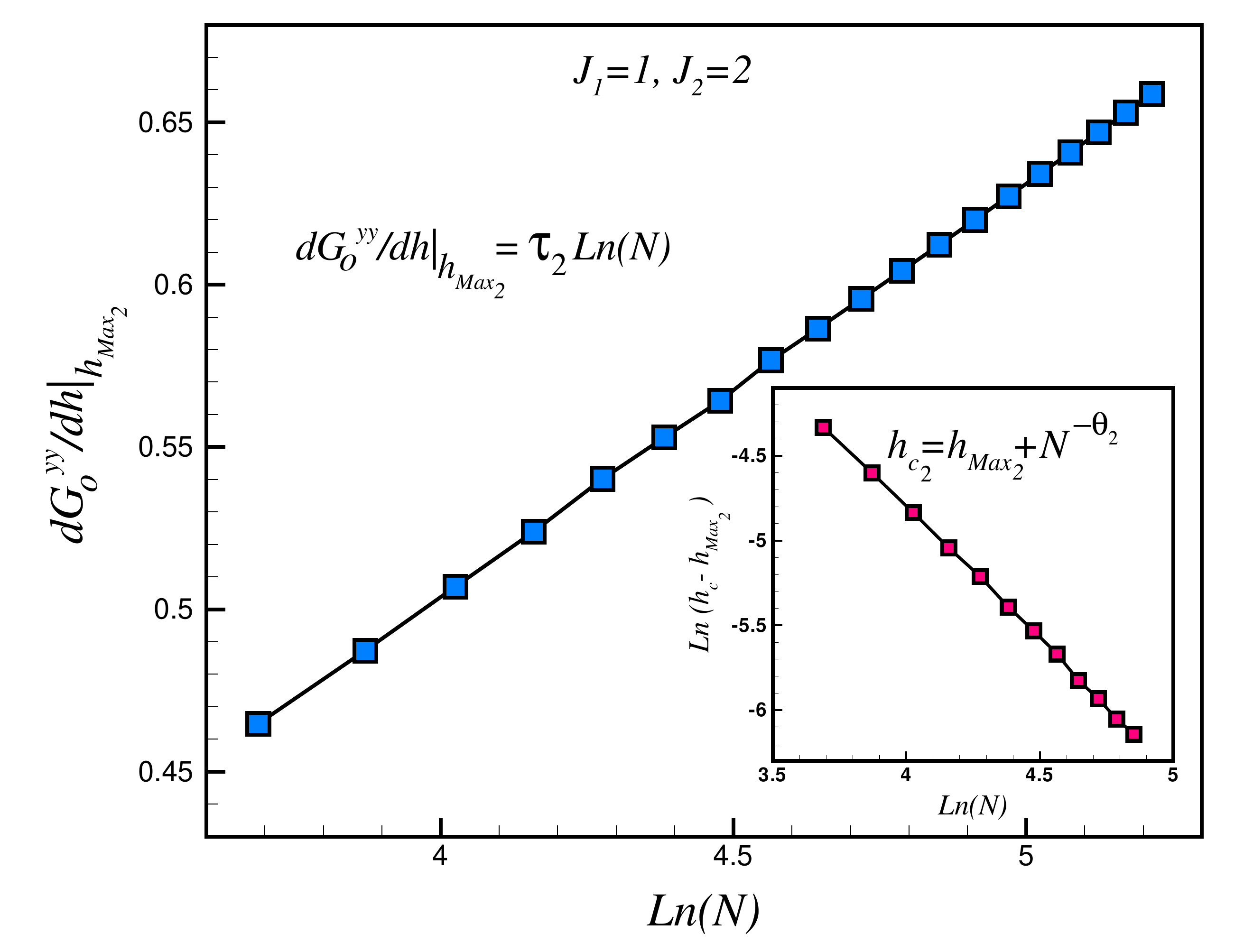}
\caption{(Color online) The logarithm of the second maximum of $\frac{dG^{yy}}{dh}|$ versus the logarithm of chain size, $\ln(N)$, which is linear and shows a scaling behavior ($J_{1}=1, J_{2}=2$). Inset: The scaling behavior of $h_{Max_{2}}$ in terms of system size ($N$) where $h_{Max_{2}}$ is the position of second maximum in Fig.(\ref{fig10})} \label{fig12}
\end{center}
\end{figure}
%%%%%%%%%%%%%%%%%%%%%%%%%%%%%%%%%%%%%%%%%%%%%%%%%%%%%%%

%%%%%%%%%%%%%%%%%%%%%%  Fig.13   %%%%%%%%%%%%%%%%%%%%%%%
\begin{figure}
\begin{center}
\includegraphics[width=8cm]{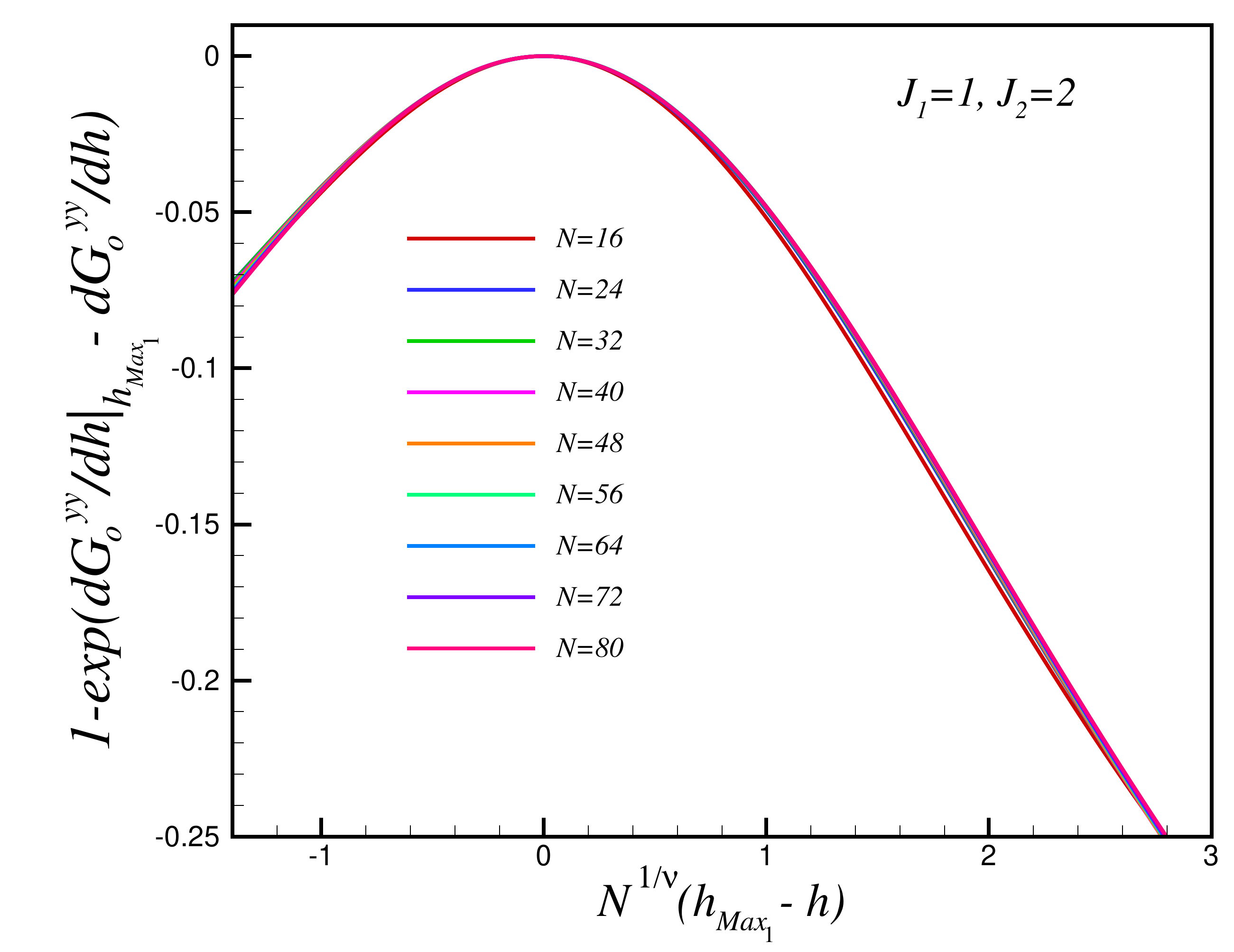}
\caption{(Color online) The finite-size scaling analysis for the
case of logarithmic divergence around the first maximum point ($h_{Max_{1}}$) for $J_{1}=1, J_{2}=2$.
The NNC function, considered as a function of system size and coupling,
collapses on a single curve for different lattice sizes.} \label{fig13}
\end{center}
\end{figure}
%%%%%%%%%%%%%%%%%%%%%%%%%%%%%%%%%%%%%%%%%%%%%%%%%%%%%%%

%%%%%%%%%%%%%%%%%%%%%%  Fig.14   %%%%%%%%%%%%%%%%%%%%%%%
\begin{figure}
\begin{center}
\includegraphics[width=9cm]{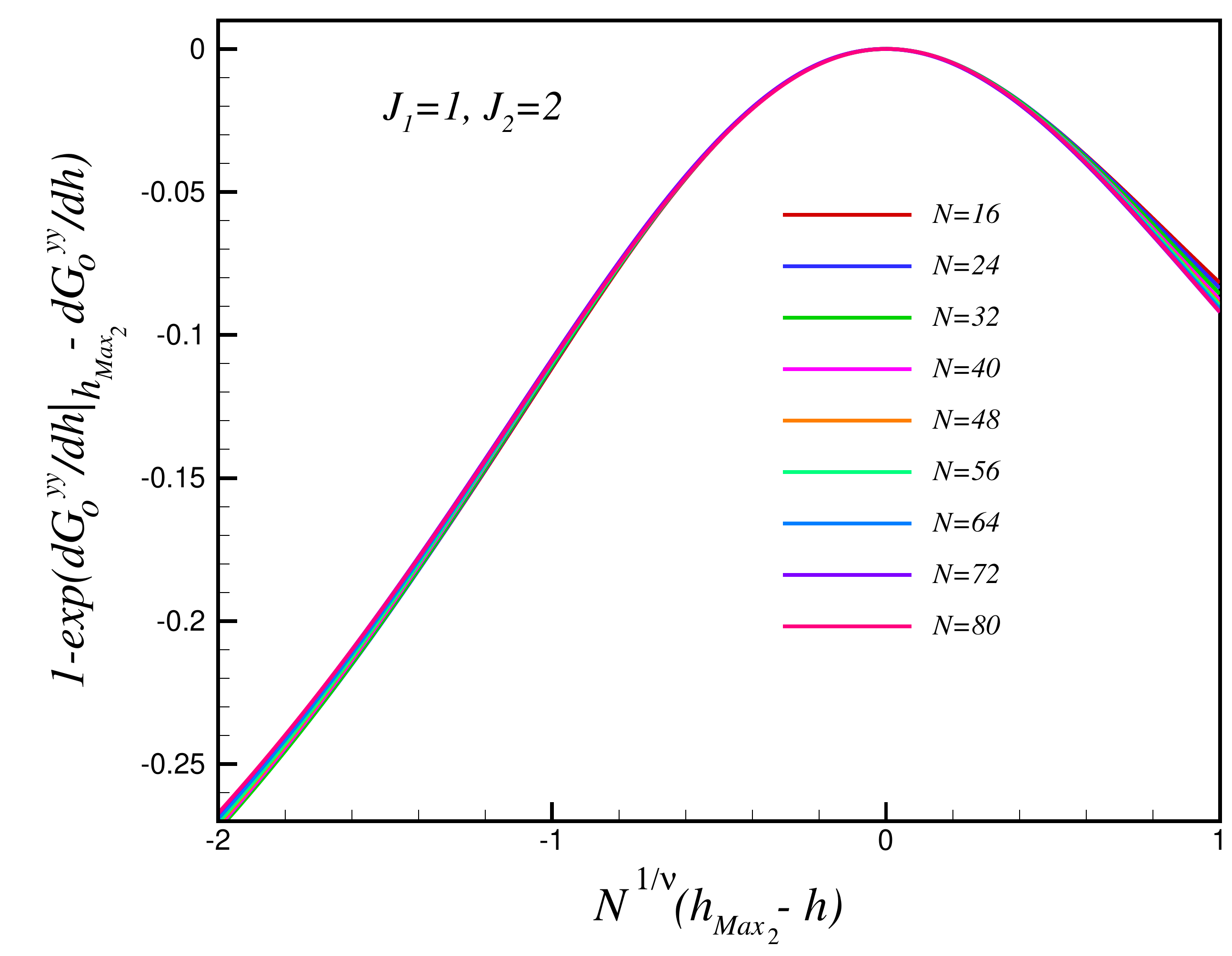}
\caption{(Color online) A manifestation of finite-size scaling of $\frac{dG^{yy}}{dh}$ around the second maximum
point ($h_{Max_{2}}$) for various system sizes in region (II).} \label{fig14}
\end{center}
\end{figure}
%%%%%%%%%%%%%%%%%%%%%%%%%%%%%%%%%%%%%%%%%%%%%%%%%%%%%%%

Moreover, we have shown that the position of the first maximum ($h_{Max{1}}$) of $dG_{o}^{yy}/dh$ goes to the first critical point, such as $h_{Max_{1}}=h_{c_{1}}-N^{-\theta_{1}}$	with $\theta_{1}=1.59\pm0.01$ (Fig. (\ref{fig11}, inset)).
The similar investigation shows the scaling behavior of $\frac{dG_{o}^{yy}}{dh}$ at the second maximum point ($h_{Max_{1}}$) where has been presented in Fig. (\ref{fig12}). It specifies a linear behavior of $\frac{dG_{o}^{yy}}{dh}|_{h_{Max_{2}}}$ versus $\ln(N)$ with the same exponent as $\frac{dG_{o}^{yy}}{dh}$ treat at the first maximum point ($\tau_{2}=0.14\pm0.02$). However, the second maximum position of correlation functions and TS show the scaling behaviors with the same exponents (Fig. (\ref{fig12}), inset) as the position of the first maximum of correlation functions and TS do.

\begin{figure}
\begin{center}
\includegraphics[width=9cm]{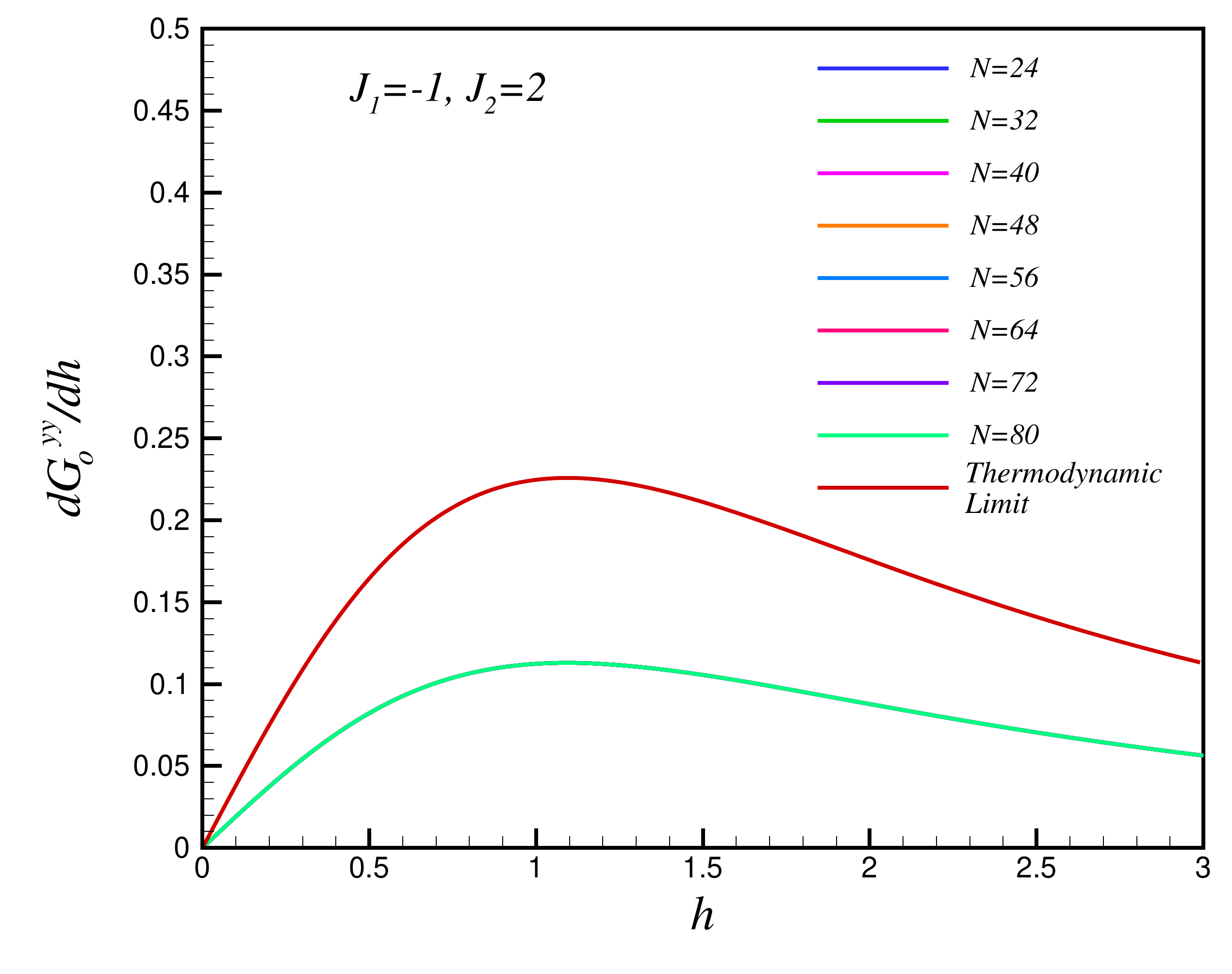}
\caption{(Color online) The derivative of $dG^{yy}$ versus
$h$ in region (III) for $J_{1}=-1, J_{2}=2$. Even in the
thermodynamic limit no singularity is observed.} \label{fig15}
\end{center}
\end{figure}

We illustrate the finite size scaling behaviors of $\frac{dG_{o}^{yy}}{dh}$ around its first and second maximum
points in figs. (\ref{fig14}) and (\ref{fig15}) respectively. They show that the NNC functions can be approximately collapsed to a single curve. These results show that all the key ingredients of
the finite size scaling are present in these cases too. In this cases scaling is fulfilled with the critical exponent $\nu=1$ in agreement with the previous results and universality hypothesis.
A similar analysis show that the universality and scaling behavior of $G_{o}^{xx}$, $G_{E}^{xx}$ and TS ($\chi^{z}$) are the same as each other in region (II).

%%%%%%%%%%%%%%%%%%%%%%  Fig.15   %%%%%%%%%%%%%%%%%%%%%%%
%%%%%%%%%%%%%%%%%%%%%%%%%%%%%%%%%%%%%%%%%%%%%%%%%%%%%%%

In Fig. (\ref{fig15}) the derivative of $G_{o}^{xx}$ with respect to magnetic field has been shown versus $h$ for $J_{1}=-1, J_{2}=2$ (region (III)). As it is clear the correlation functions does not show any singularity in this region even at the thermodynamic limit. This is justify our previous finding that shows no transition in this region. We have also investigate the TS and NNC functions behavior in the region (II). Our examination show the logarithmic divergence and scaling behavior of them close to the critical point where has been obtained using the energy gap analysis.

\section{Summary and conclusions \label{conclusion}}
In this work we have studied the quantum phase transition in the one-dimensional extended quantum compass model in the presence of a transverse magnetic field. We have shown that there are strip-antiferromagnetic, spin-flop and saturate ferromagnetic phases in addition to the phase with anti parallel ordering of spin $y$ component on odd bonds where phases separated from each other by the critical surfaces. We obtain the analytic expressions of all critical fields for the field-induced quantum phase transitions (QPT).
However we have investigate the universality and scaling properties of the nearest neighbor correlation functions and transverse susceptibility to confirm the results were obtained using the energy gap analysis. The results show that the transverse susceptibility and derivatives of the nearest neighbor correlation functions diverge close to the critical point and exhibit beautiful scaling law. So we predict that as the correlation length diverges at the critical point for an infinite system size, the derivative of correlation functions between far neighbors could capture the quantum critical point too.
The obtained exponents ($\nu=1$) and universality behaviors (logarithmic divergence) are nearly the same as those in the $1D$ transverse-field Ising model \cite{Jafari1}, suggesting that these two models share the same universality class.

Further investigations including blocks and multi-body correlations functions and the effect of temperature may
be interesting to establish a precise comparison between universality and scaling behaviors of correlation functions  and entanglement at critical points. Moreover, dynamics of correlation functions may also provide an interesting scenario for the discussion of the properties of the correlation functions and its implications for phase transitions. Such topics are left for a future research.

%%%%%%%%%%%%%%%%%%%%%%%%%%%%%%%%%%%%%%%%%%%%%%%%%%%%%%%%%%%%%%%%%%
%\begin{acknowledgments}
%The authors would like to acknowledge for useful discussions and comments.
%\end{acknowledgments}
%%%%%%%%%%%%%%%%%%%%%%%%%%%%%%%%%%%%%%%%%%%%%%%%%%%%%%%%%%%%%%%%%

\end{document}